# Ensuring Computer Science Learning in the AI Era: Open Generative AI Policies and Assignment-Driven Written Quizzes

Chan-Jin Chung
*Department of Math and Computer Science*
*Lawrence Technological University*
Southfield, Michigan, USA
cchung@LTU.edu

*Abstract*—The widespread availability of generative artificial intelligence (GenAI) has created a pressing challenge in computer science (CS) education: how to incorporate powerful AI tools into programming coursework without undermining student learning through cognitive offloading. This paper presents an assessment model that permits the use of Generative AI for take-home programming assignments while enforcing individual mastery through immediate, assignment-driven written quizzes. To promote authentic learning, these in-class, closed-book assessments are weighted more heavily than the assignments themselves and are specifically designed to verify the student's comprehension of the algorithms, structure, and implementation details of their submitted code. Preliminary empirical data were collected from an upper-level computer science course to examine the relationship between self-reported GenAI usage and performance on AI-free quizzes, exams, and final course grades. Statistical analyses revealed no meaningful linear correlation between GenAI usage levels and assessment outcomes, with Pearson correlation coefficients consistently near zero. These preliminary results suggest that allowing GenAI for programming assignments does not diminish students' mastery of course concepts when learning is verified through targeted, assignment-driven quizzes. Although limited by a small sample size, this study provides preliminary evidence that the risks of cognitive offloading can be mitigated by allowing AI-assisted programming practice while verifying understanding through assignment-driven, AI-free quizzes. The findings support the responsible adoption of open GenAI policies in upper-level CS courses, when paired with rigorous, independent assessment mechanisms.

I. INTRODUCTION

Rooted in educational constructivism, programming assignments serve as the primary engine of conceptual learning in Computer Science, bridging the gap between abstract theory and concrete practice [1]. By actively implementing and testing algorithms, students construct viable mental models, thereby solidifying their understanding of core computer science (CS) concepts.

Empirical studies have consistently shown that the hands-on cognitive effort required to implement an algorithm is what transforms algorithmic ideas into deep understanding [2].

However, the emergence of Generative AI (GenAI) has disrupted traditional Computer Science education by producing high-fidelity code for traditional 'write-from-scratch' take-home assignments [7].

Recent literature warns that the unguided use of GenAI may lead to "cognitive offloading", where learners delegate essential cognitive processing to external tools, potentially hindering deep learning [3]. In the context of programming, this offloading allows students to bypass critical stages such as algorithm design and debugging, resulting in high assignment scores but low retention of fundamental concepts. Although generative AI can transform how code is produced, the necessity of the hands-on 'theory-practice' loop for student learners remains unchanged [4]; if the student bypasses the implementation struggle entirely, the conceptual solidification fails to occur [5].

Banning GenAI is ineffective and misaligned with industry trends [6], but unrestricted use raises concerns about "hollow" or "fake" learning (students generating code they don't understand) due to cognitive offloading. Furthermore, completely banning GenAI is technically infeasible for take-home assignments.

This article proposes a new class assessment model based on "open and verify", where "open" means allowing the use of GenAI and "verify" means in-class written quizzes immediately following assignment deadlines. We assume this assessment model is for Upper-level Computer Science courses where students have already acquired basic programming skills after passing introductory programming courses such as CS1 and CS2 [8]. Research Questions in this study include:

- RQ1: Is there a correlation between the GenAI usage percentage and course grades?
- RQ2: Does allowing GenAI improve or hinder students' learning in upper-level CS courses?
- RQ3: Do assignment-driven written quizzes ensure effective learning in CS courses when AI is open to use for assignments?

The next section introduces methodology designed and used in this study, section III shows data and analysis results from a course as an empirical study, section IV discusses the findings after the analysis, and last section V summarizes findings, contributions, and future work.

## II. METHODOLOGY

Empirical data was collected from MCS 5993 Evolutionary Computation and Deep learning class being taught by the author. The course is for both upper level undergraduate and first year graduate students. Students' performance was evaluated using the model shown in Table I in previous years without officially the use of GenAI for assignment. Under the new model, individual programming assignments were weighted at 2% each, while the corresponding in-class, closed-book quizzes carried substantially higher weight, 5%. Quizzes are specifically designed to verify the student's comprehension of the algorithms, structure, and implementation details of their submitted code. Table II shows a new assessment model used in the fall 2025 semester with the following main changes:

- Use of GenAI tools is officially allowed but students must understand every line and be capable of modifying and extending the code. A disclaimer (See Fig. 1) is required for every submitted source code, Jupyter notebook file.

- Lower weight on pure programming implementation, from 5% to 2%.

- Immediately following each programming assignment deadline, students completed proctored, written verification quizzes. Questions are directly related to the programming assignment. Examples are "Describe your algorithm implemented for your assignment in pseudo code", "Complete missing lines of a sample code", "Explain lines 40-50", and "Write the output of a given code".

TABLE I. OLD ASSESSMENT MODEL IN PREVIOUS YEARS

| take-home programming assignments 1-4 (5% each) | 20% |
|---|---|
| midterm exam | 15% |
| take-home programming assignments 5, 6 | 10% |
| term project | 25% |
| final exam | 30% |

TABLE II. NEW ASSESSMENT MODEL IN FALL 2025

| take-home programming assignments 1 | 2% |
|---|---|
| written quiz #1 | 5% |
| take-home programming assignments 2, 3 | 4% |
| written quiz #2 | 10% |
| take-home programming assignments 4, 5 | 4% |
| written quiz #3 | 10% |
| take-home programming assignments 6 | 7% |
| term project | 28% |
| final exam (a part was quiz #4 from assignment 6) | 30% |

**Take-home Assignment Disclaimer**
- **Author Name:** ______________ (LTU ID: ____________)
- **Work Ownership:** This work is my own. It is not a copy from someone (Yes/No): ___
- **GenAI Assistance:** Percentage of code generated with AI tools: ___% (If greater than 0%, briefly describe how AI was used: _________________________________)
- **Understanding:** I understand every part of this code (Yes/No): ___
- **Confidence:** I am confident that I can modify, adapt, and extend this code on my own (Yes/No): ___

Fig. 1. A disclaimer template required for every source code file

In the fall 2025 semester at Lawrence Technological University (LTU), the course had 14 (13 graduate and 1 undergraduate) students. The following data was collected with the permission of LTU's IRB under Exemption Category 2, Approval #02725. Each student signed an Informed Consent form and grades were not affected by survey participation.

- Self-reported AI usage percentage for each assignment vs. corresponding quiz scores
- Self-reported overall AI usage percentage vs. Final course grades
- End-of-semester efficacy survey (Multiple Choice or Multiple Ans Questions)
- End-of-semester efficacy survey (general comments)

## III. RESULTS

### A. Correlation between the AI usage percentages and assignment scores & quizzes

Fig. 2 illustrates the relationship between students' HW1 scores (out of 2.0) and their self-reported GenAI usage. Sample size is 14. The dashed line depicts the linear regression analysis of these variables. The line is the "best guess" for the relationship based on the specific data points. The Pearson correlation coefficient ($r = 0.090$) suggests a negligible relationship. Furthermore, the large p-value ($0.761 > 0.05$) indicates that this correlation is not statistically significant. This reinforces the conclusion that there is no meaningful linear association between AI usage and the assignment scores. The light gray shaded area around the dashed line represents the 95% Confidence Interval. The shaded area shows the range of where the "true" line likely falls. Narrow area means we are very confident about the trend (usually happens with lots of data or a very clear pattern). A wide area means there is a lot of uncertainty. In summary, the wide shaded area visually confirms that AI use percentages are not a reliable predictor for HW1 scores in this dataset.

Right after the first take-home assignment (HW1) deadline, the first in-person written quiz was conducted. HW1 was designed to check students' basic programming skills in Python and understanding of gradient based function optimization algorithms. Fig. 3 shows the relationship between students' Quiz 1 scores (out of 5.0) and their self-reported GenAI usage. The dashed line depicts the linear relationship of these variables. The Pearson correlation coefficient ($r = 0.012$) suggests a negligible relationship. Furthermore, the large p-value (0.967 > 0.05) indicates that this correlation is not statistically significant. This reinforces the conclusion that there is no meaningful linear association between AI usage and the HW1-based Quiz 1 scores.

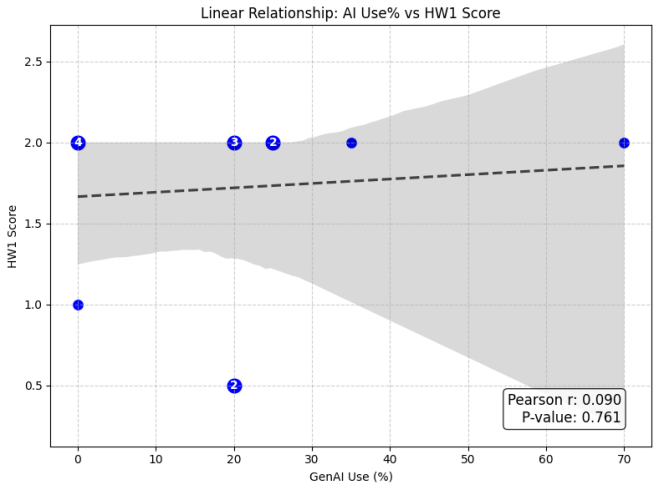

Fig. 2. Scatter plot showing the linear relationship between GenAI Use % and HW1 Score

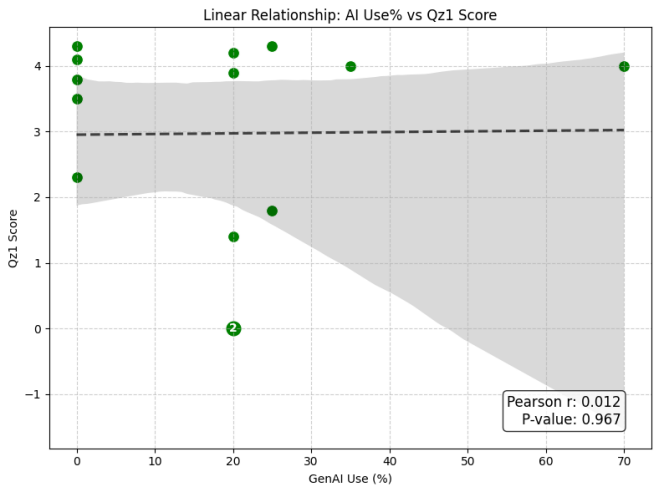

Fig. 3. Scatter plot showing the linear relationship between GenAI Use % and HW1-driven Quiz 1 Score

Table III shows the results of the relationships between students' scores and their self-reported GenAI usages for all course assignments. All nine analyses report very weak (negligible) or weak relationships. Furthermore, the large p-values ( > 0.05) indicate that the correlations are not statistically significant. This reinforces the conclusion that there is no meaningful linear association between GenAI usage and students' course scores.

Fig. 4 depicts a scatter plot showing the linear relationship between overall GenAI Use % from the end-of-semester survey and calculated total course scores for letter grades. This overview graph strengthens the finding that there is no meaningful linear association between GenAI usage and students' course scores.

TABLE III. RELATIONSHIPS BETWEEN STUDENTS' SCORES AND THEIR SELF-REPORTED GenAI USAGES FOR ALL COURSE ASSIGNMENTS.

| Relationship category | | Pearson r | *Strength of relationship** | p |
|---|---|---|---|---|
| HW 1 | GenAI use % vs. HW1 score | 0.090 | Very weak | 0.761 |
| | GenAI use % vs. Quiz1 score | 0.012 | Very weak | 0.967 |
| HW 2 and 3** | Avg GenAI use % vs. HW2_3 avg. score | 0.201 | Weak | 0.490 |
| | Avg GenAI use % vs. Quiz2 score | -0.161 | Very weak | 0.582 |
| HW 4 and 5** | Avg GenAI use % vs. HW4_5 avg. score | 0.174 | Very weak | 0.608 |
| | Avg GenAI use % vs. Quiz3 score | 0.105 | Very weak | 0.760 |
| HW 6 | GenAI use % vs. HW6 score | -0.088 | Very weak | 0.765 |
| | GenAI use % vs. Quiz4 (Part of final exam) score | 0.128 | Very weak | 0.664 |
| Total class | GenAI use % vs. Total class score | 0.105 | Very weak | 0.720 |

(*) Strength of Relationship classification rule: 0.00 to 0.19: Very Weak (often negligible), 0.20 to 0.39: Weak

(**) Outliers removed. 2 students did not take the Quizzes and received zero scores.

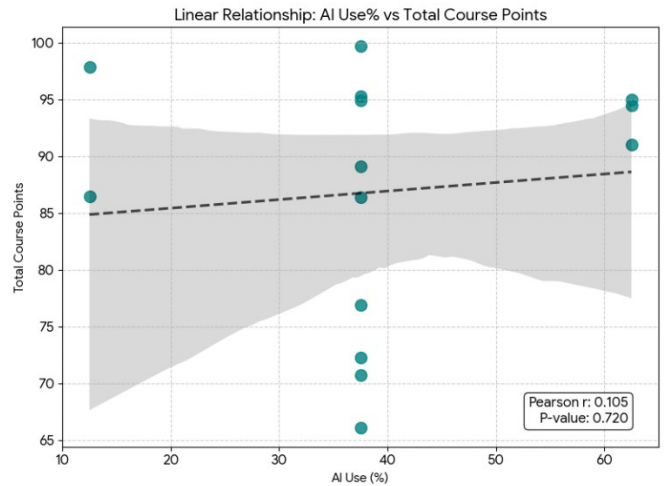

Fig. 4. Scatter plot showing the linear relationship between overall GenAI Use % and Total course scores

*B. End-of-semester Survey Results*

At the end of the semester after the final exam, students were asked to take an online survey using Canvas CMS. The survey had 7 multiple choice, 2 multiple answer questions, and one essay question. Each survey question and its corresponding result are listed below. In summary, student feedback was generally positive, with the consensus that the class assessment policy effectively ensured Computer Science learning.

**SQ1. Which generative AI tools did you use for programming assignments in this course?** (Select all that apply)

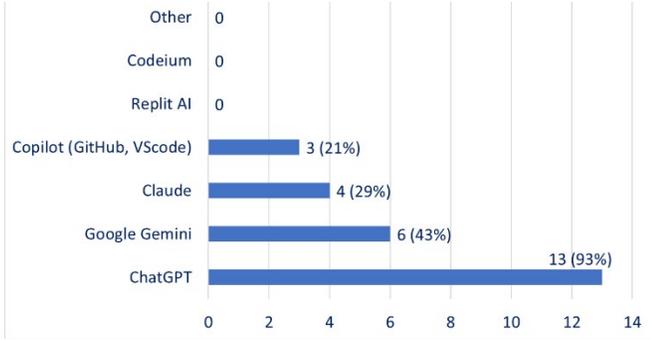

**SQ2. In what ways did you use generative AI tools? (Select all that apply)**

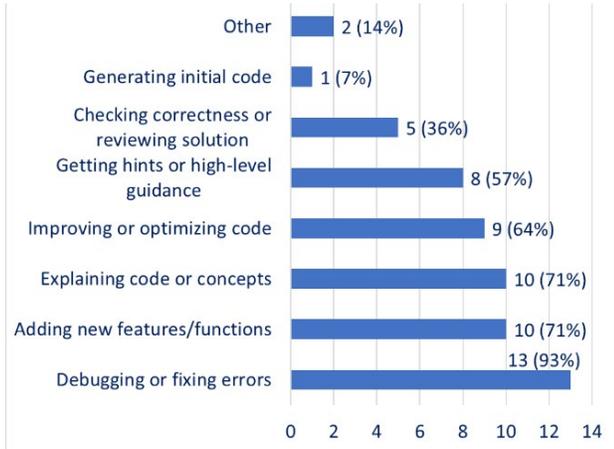

**SQ3. The written quizzes based on the assignments motivated me to read and understand the code.**

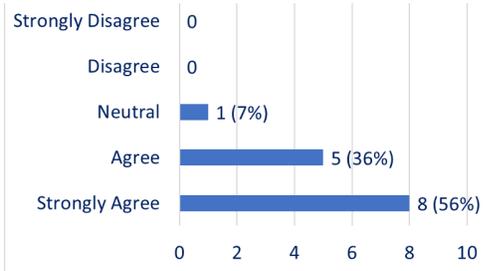

**SQ4. Approximately what percentage of your programming assignments were assisted by generative AI?**

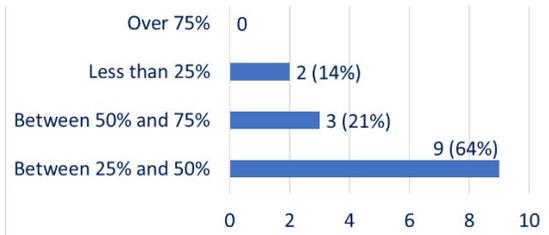

**SQ5. Allowing Generative AI reduced my stress and helped me learn more effectively.**

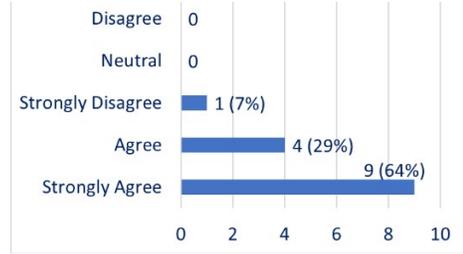

**SQ6. The programming assignments helped me understand the concepts, even though AI tools were allowed.**

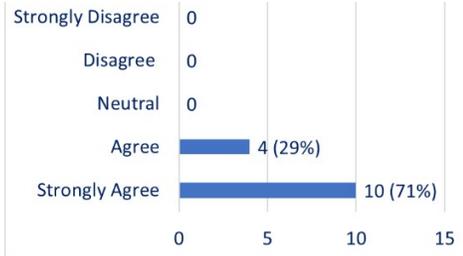

**SQ7. The written quizzes accurately tested whether I understood the assignment material.**

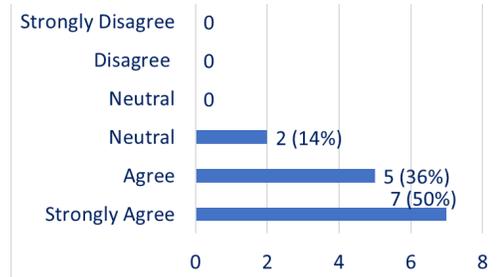

**SQ8. Overall, this assignment allowing AI + quiz strategy helped me learn the course material more effectively.**

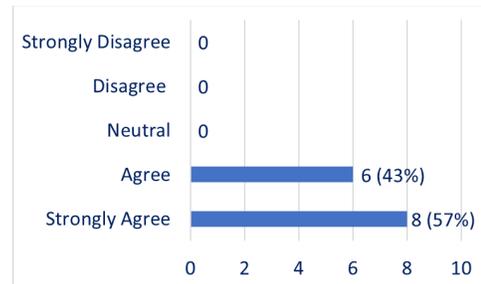

**SQ9. The class policy helped me learn how to use AI tools responsibly.**

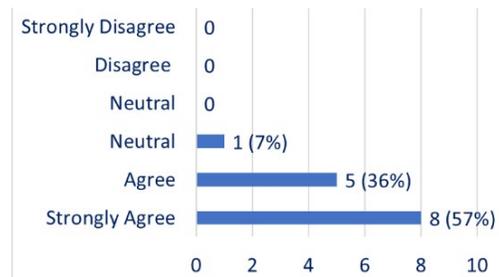

**SQ10. What changes, if any, would you recommend to improve this approach? (Not required).**

- *In the quiz if we were given an algorithm code and we were to describe in very detail what everything does and how overall the algorithm or model works I think that would have been a good addition, as I think with generative AI we will spend more time reading code than writing. (My GenAI usage: 50-75%)*

- *In my opinion, AI tools help learn concepts very well, but they also reduce the ability to complete course work without it. (My GenAI usage: less than 25%)*

- *I would recommend having more quizzes, I found them much more of a knowledge check. (25-50%)*

- *I am fine with the overall structure of course , assignments and materials provided (less than 25%)*

- *If any student uses AI for creating and understanding code, it is completely fine. But the one who can create the description of the generated code are the real ones who understood it. (25-50%)*

- *Nothing to improve. Everything is perfect. I have learned the basic use of AI and how we can use and train. (25-50%)*

- *All is good but could have learned better without restrictions on AI. Because of AI sometimes I was hesitant to explore my ideas instead used AI for better competition among peers. So, I think that I was pulled back a bit but overall it is a good stress free experience. (25-50%)*

- *no changes required, i think its perfect execution between modern AI helping and professors teaching. (25-50%)*

- *I think that written quizzes in pseudo-code may be more effective at evaluating if students understood concepts, while sidestepping concerns about syntax or familiarity with a particular language or library. (25-50%)*

- *Overall, AI usage is extremely good. (50-75%)*

- *I probably would like more of the class competitions Like Hw6 instead of usual assignments on Quizzes (25-50%)*

## IV. DISCUSSION

The data in Section III.A demonstrates no significant correlation between the percentage of GenAI usage and overall course grades, providing no evidence of a meaningful relationship with respect to Research Question 1 (**RQ1**) defined in Section I. These findings suggest that when mandatory, assessment-driven quizzes are used to verify mastery, a student's course performance remains grounded in their individual ability to solve Computer Science problems rather than their reliance on GenAI tools.

Concerning RQ2, based on the end-of-semester Survey Results in section III.B, allowing GenAI improved students' learning in upper-level CS courses and did not hinder the learning when assignment-driven written quizzes are conducted and administered. Specifically, all the students strongly agreed or agreed that the programming assignments helped them understand the concepts, even though AI tools were allowed (see SQ6). In SQ9 all the students strongly agreed or agreed the assignment allowing AI plus immediate quiz strategy helped them learn the course material more effectively. 93% of students said allowing Generative AI reduced their stress and helped them learn more effectively (see SQ5).

Regarding RQ3, assignment-driven verification quizzes ensure effective learning in CS courses when AI is open to use for assignments based on the following survey results:

- Students commented on how the quizzes forced them to read/debug the AI generated code rather than blindly submitting it. 93% of students said: The written quizzes based on the assignments motivated me to read and understand the code before submitting.

- 100% of students said: The programming assignments helped me understand the concepts, even though AI tools were allowed.

- 93% of students said: The class policy helped me learn how to use AI tools responsibly.

- 86% of students said: The written quizzes accurately tested whether I understood the assignment material.

- The quizzes appear successfully to mitigate the risk of plagiarism by forcing "Code Review" habits using the AI tools. 71% of students said AI tools were used to explain code or concepts (see SQ2).

Fig. 5 shows Average Total Course Points for the 3 groups (High AI users, Moderate AI users, and Low AI users) based on SQ4. The distribution of total points by AI usage is also plotted in Figure 4. Interestingly, the students with moderate AI usage (25-50%) had the lowest average scores, while those with very low (<25%) and very high (50-75%) usage performed better. The moderate AI use group *may* lack the confidence to write manual code but also lack the "Prompt Engineering" skills to employ the AI effectively.

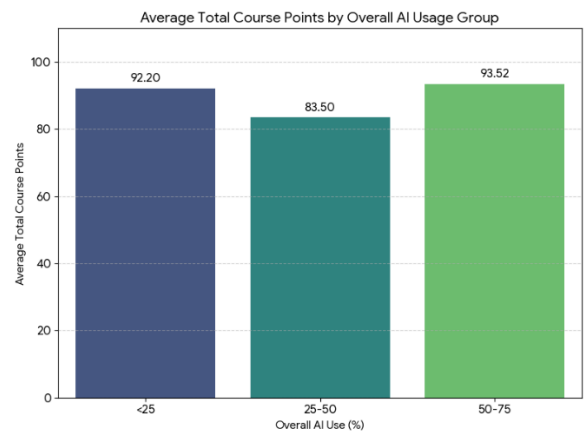

Fig. 5. Average Total Course Points by Overall AI Usage Groups

## V. SUMMARY OF FINDINGS AND CONCLUSION

The Pearson correlation coefficient between self-reported generative AI usage and final course points was $r \approx 0.11$, indicating a very weak positive linear relationship when the open but verify model is adopted. This result suggests that AI usage level alone was not a strong predictor of student performance under the adopted assessment structure in the class. In other words, the assessment-driven quizzes based on the open-but-verify concept mitigated the risks of cognitive offloading even if GenAI is open in the class. The only disadvantage of this approach is additional work for the instructor due to more exams. While the results are promising, the small sample size (*n*=14) means they should be treated as exploratory. Furthermore, reliance on self-reported data and the dual role of the instructor as researcher may have introduced response bias.

Future Work includes expanding the sample size and analyzing how the high-performing group prompted the AI (analyzing prompt logs), possibly personalizing quizzes by using AI to generate personalized quiz questions. For quizzes, instead of traditional written exam style questions, oral exams may be more effective to evaluate their knowledge and reduce inappropriate use of AI tools.